\documentclass[11pt]{article}
\usepackage{amsmath,amssymb,latexsym}

\advance \topmargin by -\headheight
\advance \topmargin by -\headsep
     
\evensidemargin \oddsidemargin
\newcommand\rf[1]{(\ref{eq:#1})}
\newcommand\lab[1]{\label{eq:#1}}
\newcommand\nonu{\nonumber}
\newcommand\br{\begin{eqnarray}}
\newcommand\er{\end{eqnarray}}
\newcommand\be{\begin{equation}}
\newcommand\ee{\end{equation}}

\newcommand\foot[1]{\footnotemark\footnotetext{#1}}
\newcommand\lb{\lbrack}
\newcommand\rb{\rbrack}

\newcommand\llb{\left\lbrack}
\newcommand\rrb{\right\rbrack}

\renewcommand\({\left(}
\renewcommand\){\right)}
\renewcommand\v{\vert}                     
\newcommand\bgv{\bigg\vert}              

\newcommand\bc{\begin{center}}
\newcommand\ec{\end{center}}

\newcommand\Tr{\mathop{\mathrm Tr}}                  
\newcommand\partder[2]{\frac{{\partial {#1}}}{{\partial {#2}}}}

\renewcommand\a{\alpha}

\renewcommand\d{\delta}

\newcommand\vareps{\varepsilon}
\newcommand\g{\gamma}
\newcommand\G{\Gamma}

\newcommand\h{\frac{1}{2}}
\renewcommand\k{\kappa}
\renewcommand\l{\lambda}
\renewcommand\L{\Lambda}
\newcommand\m{\mu}
\newcommand\n{\nu}
\renewcommand\o{\over}

\newcommand\p{\phi}
\newcommand\vp{\varphi}
\renewcommand\P{\Phi}
\newcommand\pa{\partial}

\newcommand\pr{\prime}

\renewcommand\r{\rho}
\newcommand\s{\sigma}

\renewcommand\t{\tau}
\renewcommand\th{\theta}

\newcommand\wti{\widetilde}
\newcommand\twomat[4]{\left(\begin{array}{cc}  
{#1} & {#2} \\ {#3} & {#4} \end{array} \right)}

\newcommand\cA{{\mathcal A}}

\newcommand\cF{{\mathcal F}}

\newcommand\cL{{\mathcal L}}

\newcommand{\ct}[1]{\cite{#1}}
\newcommand{\bib}[1]{\bibitem{#1}}
\newcommand\NPB[3]{(#2), \textsl{Nucl. Phys.} \textbf{B#1} #3}

\newcommand\PRD[3]{(#2), \textsl{Phys. Rev.} \textbf{D#1} #3}

\newcommand\PLB[3]{(#2), \textsl{Phys. Lett.} \textbf{#1B} #3}
\newcommand\CQG[3]{(#2), \textsl{Class. Quantum Grav.} \textbf{#1} #3}

\begin{document}

\title{Weyl-Invariant Light-Like Branes \\
and Black Hole Physics${}$\foot{Based on talks delivered at the
2nd Workshop ``Gravity, Astrophysics and Strings'', Kiten (Bulgaria), the 3rd 
Summer School on Modern Mathematical Physics, Zlatibor (Serbia and Montenegro), 2004,
and the 2nd Annual Meeting of the European RTN \textsl{EUCLID}, Sozopol (Bulgaria),
2004.}}

\author{Eduardo Guendelman and Alexander Kaganovich\\
\small\it Department of Physics, Ben-Gurion University, Beer-Sheva, Israel  \\[-1.mm]
\small\it email: guendel@bgumail.bgu.ac.il, alexk@bgumail.bgu.ac.il \\
${}$ \\
Emil Nissimov and Svetlana Pacheva\\
\small\it Institute for Nuclear Research and Nuclear Energy,\\[-1.mm]
\small\it Bulgarian Academy of Sciences, Sofia, Bulgaria  \\[-1.mm]
\small\it email: nissimov@inrne.bas.bg, svetlana@inrne.bas.bg}
\date{ }
\maketitle

\begin{abstract}
We propose a new class of $p$-brane theories which are Weyl-conformally invariant
for any p. For any odd world-volume dimension the latter describe intrinsically
light-like branes, hence the name \textsl{WILL}-branes (Weyl-Invariant Light-Like 
branes). Next we discuss the dynamics of \textsl{WILL}-membranes (i.e., for $p=2$)
both as test branes in various external physically relevant $D=4$ gravitational 
backgrounds, as well as within the framework of a coupled $D=4$ 
Einstein-Maxwell-\textsl{WILL}-membrane system. In all cases we find that 
the \textsl{WILL}-membrane materializes the event horizon of the corresponding black
hole solutions, thus providing an explicit dynamical realization of the
membrane paradigm in black hole physics.
\end{abstract}   

\section{Introduction - Main Motivation}   

The consistent Lagrangian formulation of geometrically motivated field theories 
(gravity, strings, branes, \textsl{etc.}; for a background on string and 
brane theories, see refs.\ct{brane-string-rev}.) requires among other things
reparametrization-covariant (generally-covariant) integration measure
densities (volume-forms). The usual choice is the standard Riemannian
integration measure given by $\sqrt{-g}$ with $g \equiv \det\v\v
g_{\m\n}\v\v$, where $g_{\m\n}$ indicates the intrinsic Riemannian metric on
the underlying manifold. 

However, equally well-suited is the following alternative non-Riemannian
integration measure density:
\be
\P (\vp) \equiv {1\o {D!}}\vareps^{\m_1 \ldots \m_D}
\vareps_{i_1 \ldots i_D} \pa_{\m_1} \vp^{i_1} \ldots \pa_{\m_D} \vp^{i_D}
\quad ,\quad i=1,\ldots, D
\lab{mod-measure-D}
\ee
built in terms of $D$ auxiliary scalar fields independent of the intrinsic
Riemannian metric.

In a series of papers \ct{TMT-basic} two of us have proposed new classes of
models involving gravity, called \textsl{two-measure theories}, 
whose actions contain both standard Riemannian {\em and}~ alternative 
non-Riemannian integration measures :
\br
S = \int d^D x\, \P (\vp)\, L_1 + \int d^D x\,\sqrt{-g}\, L_2
\nonu
\er
The scalar Lagrangians are of the following generic form:
\br
L_1 = e^{\frac{\a \phi}{M_P}} \Bigl\lb - {1\o \k} R(g,\G) -
\h g^{\m\n}\pa_\m \phi \pa_\n \phi
+ \bigl(\mathrm{Higgs}\bigr) + \bigl(\mathrm{fermions}\bigr)\Bigr\rb
\nonu
\er
and similarly for $L_2$ (with different choice of the normalization factors
in front of each of the terms).
Here $R(g,\G)$ is the scalar curvature in the first order formalism, $\phi$
is the dilaton field, $M_P$ denotes the Planck mass, \textsl{etc}.
The auxiliary fields $\vp^i$ are pure-gauge degrees of freedom except for
the new dynamical ``geometric'' field $\zeta (x) \equiv \frac{\P (\vp)}{\sqrt{-g}}$,
whose dynamics is determined only through the matter fields locally (\textsl{i.e.},
without gravitational interaction).

Two-measure theories address various basic problems in cosmology and
particle physics, and provide plausible solutions for a broad array of issues, 
such as:
\begin{itemize}
\item
Scale invariance and its dynamical breakdown; Spontaneous generation of
dimensionfull fundamental scales;  
\item
Cosmological constant problem;  
\item
The problem of fermionic families; 
\item
Applications in modern brane-world scenarios.
\end{itemize}  
For a detailed exposition we refer to the series of papers \ct{TMT-basic,TMT-recent}.

Subsequently, the idea of employing an alternative non-Riemannian integration
measure was applied systematically to string, $p$-brane and $Dp$-brane models
\ct{m-string}. The main feature of these new classes of modified
string/brane theories is the appearance of the pertinent string/brane
tension as an additional dynamical degree of freedom beyond the usual string/brane
physical degrees of freedom, instead of being introduced \textsl{ad hoc} as
a dimensionfull scale. In the next section we briefly recall the
construction of the modified bosonic string model with a dynamical tension
before proceeding to our main task. It is the construction of a novel class of
$p$-brane theories which are Weyl-conformal invariant for any $p$~ and whose
dynamics significantly differs both from the standard Nambu-Goto (or
Dirac-Born-Infeld) branes as well as from their modified versions with
dynamical string/brane tensions \ct{m-string} mentioned above. 

\section{Strings and Branes with a Modified World-Sheet/World-Volume
Integration Measure}

The modified-measure bosonic string model is given by the following action:
\br
S = - \int d^2\s\,\P (\vp) \Bigl\lb \h\g^{ab} \pa_a X^{\m} \pa_b X^{\n}G_{\m\n}(X)
- \frac{\vareps^{ab}}{2\sqrt{-\g}} F_{ab}(A)\Bigr\rb 
\nonu \\
+ \int d^2\s\,\sqrt{-\g} A_a J^a  \qquad; \quad 
J^a = \frac{\vareps^{ab}}{\sqrt{-\g}}\pa_b u \; ,
\lab{m-string}
\er
with the notations:
\be
\P (\vp) \equiv \h \vareps_{ij} \vareps^{ab} \pa_a \vp^i \pa_b \vp^j
\quad ,\quad F_{ab} (A) = \pa_{a} A_{b} - \pa_{b} A_{a} \; ,
\lab{m-string-notat}
\ee
$\g_{ab}$ denotes the intrinsic Riemannian world-sheet metric with 
$\g = \det\Vert\g_{ab}\Vert$~ and $G_{\m\n}(X)$ is the Riemannian metric of
the embedding space-time ($a,b=0,1; i,j=1,2; \m,\n =0,1,\ldots,D-1$).

Here is the list of differences w.r.t. the standard Nambu-Goto string
(in the Polyakov-like formulation) :
\begin{itemize}
\item
New non-Riemannian integration measure density $\P (\vp)$ instead of $\sqrt{-\g}$;
\item
Dynamical string tension $T \equiv \frac{\P (\vp)}{\sqrt{-\g}}$ instead of
\textsl{ad hoc} dimensionfull constant;
\item
Auxiliary world-sheet gauge field $A_a$ in a would-be ``topological'' term
$\int d^2\s\, \frac{\P (\vp)}{\sqrt{-\g}} \h\vareps^{ab} F_{ab}(A)$;
\item
Optional natural coupling of auxiliary $A_a$ to external conserved world-sheet 
electric current $J^a$ (see last equality in \rf{m-string} and 
Eq.\rf{Maxwell-like-eqs} below).
\end{itemize}

The modified string model \rf{m-string} is Weyl-conformally invariant
similarly to the ordinary case. Here Weyl-conformal symmetry is given by 
Weyl rescaling of $\g_{ab}$ supplemented with a special diffeomorphism in 
$\vp$-target space:
\be
\g_{ab} \longrightarrow \g^{\pr}_{ab} = \rho\,\g_{ab}  \quad ,\quad
\vp^{i} \longrightarrow \vp^{\pr\, i} = \vp^{\pr\, i} (\vp) 
\;\; \mathrm{with} \;\; 
\det \Bigl\Vert \frac{\pa\vp^{\pr\, i}}{\pa\vp^j} \Bigr\Vert = \rho \; .
\lab{Weyl-conf}
\ee

The dynamical string tension appears as a canonically conjugated momentum
w.r.t. $A_1$: 
$\pi_{A_1} \equiv \partder{\cL}{\dot{A_1}} = \frac{\P (\vp)}{\sqrt{-\g}}
\equiv T$, \textsl{i.e.}, $T$ has the meaning of a 
\textit{world-sheet electric field strength}, and the eqs. of motion w.r.t. 
auxiliary gauge field $A_a$ look exactly as $D=2$ Maxwell eqs.:
\be
\frac{\vareps^{ab}}{\sqrt{-\g}}\pa_b T + J^a = 0 \; .
\lab{Maxwell-like-eqs}
\ee
In particular, for $J^a= 0$ :
\be
\vareps^{ab} \pa_{b} \Bigl(\frac{\P (\vp)}{\sqrt{-\g}}\Bigr) = 0 \qquad,\quad
\frac {\P (\vp)}{\sqrt{-\gamma}}  \equiv T = \textrm{const}\; ,
\lab{Maxwell-like-eqs-0}
\ee
one gets a {\em spontaneously induced} constant string tension.
Furthermore, when the modified string couples to point-like charges on the 
world-sheet (\textsl{i.e.}, $J^0 {\sqrt{-\gamma}} = \sum_i e_i \d (\s - \s_i)$
in \rf{Maxwell-like-eqs}) one obtains classical charge {\em confinement}: 
$\sum_i e_i = 0$.

The above charge confinement mechanism has also been generalized in \ct{m-string}
to the case of coupling the modified string model with dynamical tension to 
non-Abelian world-sheet ``color'' charges. The latter is achieved as follows.
Notice the following identity in $2D$ involving Abelian gauge field $A_a$:
\be
\frac{\vareps^{ab}}{2\sqrt{-\g}} F_{ab}(A) = 
\sqrt{-\h F_{ab}(A) F_{cd}(A) \g^{ac}\g^{bd}} \; .
\lab{2D-id}
\ee 
Then the extension of the action \rf{m-string} to the non-Abelian case is
straightforward:
\be
S = - \int d^2\s \,\P (\vp) \Bigl\lb \h \g^{ab} \pa_a X^{\m} \pa_b X^\n
G_{\m\n}(X) - \sqrt{-\h \Tr (F_{ab}(A)F_{cd}(A)) \g^{ac}\g^{bd}}\Bigr\rb
+ \int d^2\s\,\Tr \( A_a j^a\)
\lab{m-string-NA}
\ee
with $F_{ab}(A) = \pa_a A_b - \pa_b A_c + i \bigl\lb A_a,\, A_b\bigr\rb$,
sharing the same principal property -- dynamical generation of string
tension as an additional degree of freedom.


\section{New Class of Weyl-Invariant $p$-Brane Theories}

\subsection{Weyl-Invariant Branes: Action and Equations of Motion}

The identity \rf{2D-id} suggests how to construct \textbf{Weyl-invariant} 
$p$-brane models for any $p$. Namely, we propose the following novel
$p$-brane actions:
\be
S = - \int d^{p+1}\s \,\P (\vp) 
\Bigl\lb \h \g^{ab} \pa_a X^{\m} \pa_b X^{\n} G_{\m\n}(X)
- \sqrt{F_{ab}(A) F_{cd}(A) \g^{ac}\g^{bd}}\Bigr\rb
\lab{WI-brane}
\ee
\be
\P (\vp) \equiv \frac{1}{(p+1)!} \vareps_{i_1\ldots i_{p+1}} 
\vareps^{a_1\ldots a_{p+1}} \pa_{a_1} \vp^{i_1}\ldots \pa_{a_{p+1}} \vp^{i_{p+1}}
\; ,
\lab{mod-measure-p}
\ee
where notations similar to those in \rf{m-string} are used 
(here $a,b=0,1,\ldots,p; i,j=1,\ldots,p+1$).

The above action \rf{WI-brane} is invariant under Weyl-conformal symmetry 
(the same as in the dynamical-tension string model \rf{m-string}):
\be
\g_{ab} \longrightarrow \g^{\pr}_{ab} = \rho\,\g_{ab}  \quad ,\quad
\vp^{i} \longrightarrow \vp^{\pr\, i} = \vp^{\pr\, i} (\vp) 
\;\; \mathrm{with} \;\; 
\det \Bigl\Vert \frac{\pa\vp^{\pr\, i}}{\pa\vp^j} \Bigr\Vert = \rho \; .
\lab{Weyl-conf-p}
\ee
  
We notice the following significant differences of \rf{WI-brane} w.r.t. the standard
Nambu-Goto $p$-branes (in the Polyakov-like formulation) :
\begin{itemize}
\item
New non-Riemannian integration measure density $\P (\vp)$ instead of $\sqrt{-\g}$,
and {\em no}~ ``cosmological-constant'' term ($(p-1)\sqrt{-\g}$);
\item
Variable brane tension $\chi \equiv \frac{\P (\vp)}{\sqrt{-\g}}$ 
which is Weyl-conformal {\em gauge dependent}: $ \chi \to \rho^{\h(1-p)}\chi$;
\item
Auxiliary world-volume gauge field $A_a$ in a ``square-root'' Maxwell
(Yang-Mills) term\foot{``Square-root'' Maxwell (Yang-Mills) action in $D=4$
was originally introduced in the first ref.\ct{Spallucci} and later
generalized to ``square-root'' actions of higher-rank
antisymmetric tensor gauge fields in $D\geq 4$ in the second and third
refs.\ct{Spallucci}.};
the latter is straightforwardly generalized to the non-Abelian case 
-- $\sqrt{-\Tr \( F_{ab}(A) F_{cd}(A)\) \g^{ac}\g^{bd}}$ similarly to
\rf{m-string-NA};
\item
Natural optional couplings of the auxiliary gauge field $A_a$ to external 
world-volume ``color'' charge currents $j^a$;
\item
The action \rf{WI-brane} is manifestly Weyl-conformal invariant for {\em any} $p$;
it describes {\em intrinsically light-like} $p$-branes for any even $p$.
\end{itemize}

The eqs. of motion w.r.t. measure-building auxiliary scalars $\vp^i$ are:
\be
\h \g^{cd}\(\pa_c X \pa_d X\) - \sqrt{FF\g\g} = M \; \Bigl( = \mathrm{const}\Bigr)
\; ,
\lab{phi-eqs}
\ee
employing the short-hand notations: 
\be
\(\pa_a X \pa_b X\) \equiv \pa_a X^\m \pa_b X^\n G_{\m\n}\quad ,\quad
\sqrt{FF\g\g} \equiv \sqrt{F_{ab} F_{cd} \g^{ac}\g^{bd}} \; .
\lab{short-hand}
\ee
The eqs. of motion w.r.t. $\g^{ab}$ read:
\be
\h\(\pa_a X \pa_b X\) + \frac{F_{ac}\g^{cd} F_{db}}{\sqrt{FF\g\g}} = 0 \; ,
\lab{gamma-eqs}
\ee
and (upon taking the trace) imply $M=0$ in Eq.\rf{phi-eqs}.

Next we have the following eqs. of motion w.r.t. auxiliary gauge field $A_a$
and w.r.t. $X^\m$, respectively:
\be
\pa_b \(\frac{F_{cd}\g^{ac}\g^{bd}}{\sqrt{FF\g\g}} \P (\vp)\) = 0 \; ,
\lab{A-eqs}
\ee
\be
\pa_a \(\P (\vp) \g^{ab}\pa_b X^\m\) +
\P (\vp) \g^{ab}\pa_a X^\n \pa_b X^\l \G^\m_{\n\l} = 0 \; ,
\lab{X-eqs}
\ee
where $\G^\m_{\n\l}=\h G^{\m\k}\(\pa_\n G_{\k\l}+\pa_\l G_{\k\n}-\pa_\k G_{\n\l}\)$
is the affine connection corresponding to the external space-time metric 
$G_{\m\n}$.

\subsection{Intrinsically Light-Like Branes}

Let us consider the $\g^{ab}$-eqs. of motion \rf{gamma-eqs}.
$F_{ab}$ is an anti-symmetric $(p+1)\times (p+1)$ matrix, therefore, $F_{ab}$ is 
{\em not invertible} in any odd $(p+1)$ -- it has at least one zero-eigenvalue 
vector $V^a$ ($F_{ab}V^b = 0$). Therefore, for any odd $(p+1)$ the induced
metric:
\be
g_{ab} \equiv \(\pa_a X \pa_b X\) \equiv \pa_a X^\m \pa_b X^\n G_{\m\n}
\lab{ind-metric}
\ee
on the world-volume of the Weyl-invariant brane \rf{WI-brane} is {\em singular} 
as {\em opposed} to the ordinary Nambu-Goto brane (where the induced metric
is proportional to the intrinsic Riemannian world-volume metric):
\be
\(\pa_a X \pa_b X\) V^b = 0 \quad ,\quad \mathrm{i.e.}\;\;
\(\pa_V X \pa_V X\) = 0 \;\; ,\;\; \(\pa_{\perp} X \pa_V X\) = 0 \; ,
\lab{LL-constraints}
\ee
where $\pa_V \equiv V^a \pa_a$ and $\pa_{\perp}$ are derivates along the
tangent vectors in the complement of the tangent vector field $V^a$. 

Thus, we arrive at the following important conclusion:
every point on the world-surface of the Weyl-invariant $p$-brane \rf{WI-brane} 
(for odd $(p+1)$) moves with the speed of light in a time-evolution along the 
zero-eigenvalue vector-field $V^a$ of the world-volume electromagnetic
field-strength $F_{ab}$. Therefore, we will name \rf{WI-brane} (for odd $(p+1)$)
by the acronym {\em WILL-brane} (Weyl-Invariant Light-Like-brane) model.
  
\subsection{Dual Formulation of {\em WILL}-Branes}

The $A_a$-eqs. of motion \rf{A-eqs} can be solved in terms of $(p-2)$-form gauge 
potentials $\L_{a_1\ldots a_{p-2}}$ dual w.r.t. $A_a$. The respective
field-strengths are related as follows:
\br
F_{ab}(A)= -\frac{1}{\chi}\,\frac{\sqrt{-\g}\,\vareps_{abc_1\ldots c_{p-1}}}{2(p-1)}
\g^{c_1 d_1}\ldots \g^{c_{p-1} d_{p-1}} 
\, F_{d_1\ldots d_{p-1}}(\L) \,\g^{cd} \(\pa_c X \pa_d X\) \; ,
\lab{dual-strength-rel} \\
\chi^2 = - \frac{2}{(p-1)^2}\, \g^{a_1 b_1}\ldots \g^{a_{p-1} b_{p-1}}
F_{a_1\ldots a_{p-1}}(\L) F_{b_1\ldots b_{p-1}}(\L) \; ,
\lab{chi-2}
\er
where $\chi \equiv \frac{\P (\vp)}{\sqrt{-\g}}$ is the variable brane tension, and:
\be
F_{a_1\ldots a_{p-1}}(\L) = (p-1) \pa_{[a_1} \L_{a_2\ldots a_{p-1}]}
\lab{dual-strength}
\ee
is the $(p-1)$-form dual field-strength.

All eqs. of motion can be equivalently derived from the following {\em dual} 
{\em WILL}-brane action:
\be
S_{\mathrm{dual}} = - \h \int d^{p+1}\s\, \chi (\g,\L) \sqrt{-\g}
\g^{ab}\pa_a X^\m \pa_b X^\n G_{\m\n}
\lab{WI-brane-dual}
\ee
with $\chi (\g,\L)$ given in \rf{chi-2} above.
  
\section{Special case $p=2$: {\em WILL}-Membrane}

The {\em WILL}-membrane dual action (particular case of \rf{WI-brane-dual} for
$p=2$) reads:
\br
S_{\mathrm{dual}} = - \h \int d^3\s\, \chi (\g,u)\,\sqrt{-\g}
\g^{ab}\(\pa_a X \pa_b X\) \; ,
\lab{WILL-membrane} \\
\chi (\g,u) \equiv \sqrt{-2\g^{cd}\pa_c u \pa_d u} \; ,
\lab{chi-1}
\er
where $u$ is the dual ``gauge'' potential w.r.t. $A_a$:
\be
F_{ab}(A) = - \frac{1}{2\chi (\g,u)} \sqrt{-\g} \vareps_{abc} \g^{cd}\pa_d u\,
\g^{ef}\!\(\pa_e X \pa_f X\)  \; .
\lab{dual-strenght-rel-3}
\ee
$S_{\mathrm{dual}}$ is manifestly Weyl-invariant (under $\g_{ab} \to \rho\g_{ab}$).

The eqs. of motion w.r.t. $\g^{ab}$, $u$ (or $A_a$), and $X^\m$ read
accordingly:
\be
\(\pa_a X \pa_b X\) + \h \g^{cd}\(\pa_c X \pa_d X\) 
\(\frac{\pa_a u \pa_b u }{\g^{ef} \pa_e u \pa_f u} - \g_{ab}\) = 0 \; ,
\lab{gamma-eqs-3}
\ee
\be
\pa_a \(\,\frac{\sqrt{-\g}\g^{ab}\pa_b u}{\chi (\g,u)}\,
\g^{cd}\(\pa_c X \pa_d X\)\,\) = 0 \; ,
\lab{u-eqs}
\ee
\be
\pa_a \(\chi (\g,u)\,\sqrt{-\g} \g^{ab}\pa_b X^\m \) +
\chi (\g,u)\,\sqrt{-\g} \g^{ab}\pa_a X^\n \pa_b X^\l \G^\m_{\n\l} = 0 \; .
\lab{X-eqs-3}
\ee
The first eq. above shows that the induced metric 
$g_{ab} \equiv \(\pa_a X \pa_b X\)$ has zero-mode eigenvector $V^a =\g^{ab}\pa_b u$.

The invariance under world-volume reparametrizations allows to introduce the
following standard (synchronous) gauge-fixing conditions:
\be
\g^{0i} = 0 \;\; (i=1,2) \quad ,\quad \g^{00} = -1 \; .
\lab{gauge-fix}
\ee

In what follows we will use the ansatz for the dual ``gauge potential'':
\be
u (\t,\s^1,\s^2) = \frac{T_0}{\sqrt{2}}\t  \; ,
\lab{u-ansatz}
\ee
where $T_0$ is an arbitrary integration constant with the dimension of membrane
tension. In particular:
\be
\chi \equiv \sqrt{-2\g^{ab}\pa_a u \pa_b u} = T_0 
\lab{chi-0}
\ee
This means that we take $\t\equiv\s^0$ to be evolution parameter along the 
zero-eigenvalue vector-field of the induced metric on the brane 
($V^a = \g^{ab}\pa_b u = \mathrm{const}\,(1,0,0)$).

  
The ansatz for $u$ \rf{u-ansatz} together with the gauge choice for $\g_{ab}$
\rf{gauge-fix} brings the eqs. of motion w.r.t. $\g^{ab}$, $u$ (or $A_a$) and 
$X^\m$ in the following form 
(recall $\(\pa_a X \pa_b X\) \equiv \pa_a X^\m \pa_b X^\n G_{\m\n}$):
\be
\(\pa_0 X \pa_0 X\) = 0 \quad ,\quad \(\pa_0 X \pa_i X\) = 0  \; ,
\lab{constr-0}
\ee
\be
\(\pa_i X\pa_j X\) - \h \g_{ij} \g^{kl}\(\pa_k X\pa_l X\) = 0  \; ,
\lab{constr-vir}
\ee
(Eqs.\rf{constr-vir} look exactly like the classical (Virasoro) constraints for an
Euclidean string theory with world-sheet parameters $(\s^1,\s^2)$);
\be
\pa_0 \(\sqrt{\g_{(2)}} \g^{kl}\(\pa_k X\pa_l X\)\) = 0  \; ,
\lab{u-eqs-fix}
\ee
where $\g_{(2)} = \det\Vert \g_{ij}\Vert$ 
(the above equation is the only remnant from the $A_a$-eqs. of motion \rf{A-eqs});
\be
\Box^{(3)} X^\m + \( - \pa_0 X^\n \pa_0 X^\l + 
\g^{kl} \pa_k X^\n \pa_l X^\l \) \G^{\m}_{\n\l} = 0  \; ,
\lab{X-eqs-3-fix}
\ee
where:
\be
\Box^{(3)} \equiv 
- \frac{1}{\sqrt{\g^{(2)}}} \pa_0 \(\sqrt{\g^{(2)}} \pa_0 \) + 
\frac{1}{\sqrt{\g^{(2)}}}\pa_i \(\sqrt{\g^{(2)}} \g^{ij} \pa_j \)   \; .
\lab{box-3}
\ee

\section{{\em WILL}-Membrane Solutions in Various Gravitational Backgrounds}    

\subsection{Example: WILL-Membrane in a PP-Wave Background}
As a simplest non-trivial example let us consider in \rf{WILL-membrane} external
space-time metric $G_{\m\n}$ for plane-polarized gravitational wave
(\textsl{pp-wave}) background:
\be
(ds)^2 = - dx^{+} dx^{-} - F(x^{+},x^I)\, (dx^{+})^2 + dx^I dx^I \; ,
\lab{pp-wave}
\ee
and employ in \rf{constr-0}--\rf{box-3} the following natural ansatz for $X^\m$ 
(here $\s^0 \equiv \t$; $I=1,\ldots,D-2$) :
\be
X^{-} = \t \quad, \quad 
X^{+}=X^{+}(\t,\s^1,\s^2) \quad, \quad X^I = X^I (\s^1,\s^2) \; .
\lab{ansatz-pp-wave}
\ee
The non-zero affine connection symbols for the pp-wave metric \rf{pp-wave}
are: $\G^{-}_{++}=\pa_{+}F$, $\G^{-}_{+I}=\pa_{I}F$, $\G^{I}_{++}=\h\pa_{I}F$.

It is straightforward to show that the solution does not depend on the form of 
the pp-wave front $F(x^{+},x^I)$ and reads:
\be
X^{+}=X^{+}_0 = \mathrm{const} \quad ,\quad 
\g_{ij} = \t\!-\!\mathrm{independent}\; ;
\lab{pp-wave-sol-1}
\ee
\be
\pa_i X^I \pa_j X^I - 
\h \g_{ij} \g^{kl} \pa_k X^I \pa_l X^I = 0 \quad ,\quad
\pa_i \(\sqrt{\g^{(2)}} \g^{ij} \pa_j X^I \) = 0 
\lab{pp-wave-sol-2}
\ee
where the latter eqs. describe a string embedded in the transverse 
$(D-2)$-dimensional flat Euclidean space.

\subsection{Example: WILL-Membrane in a Schwarzschild Black Hole}

Let us consider spherically-symmetric static gravitational background:
\be
(ds)^2 = - A(r)(dt)^2 + B(r)(dr)^2 + 
r^2 \lb (d\th)^2 + \sin^2 (\th)\,(d\p)^2\rb \; .
\lab{spherical-symm-metric}
\ee
For the Schwarzschild black hole we have $A(r) = B^{-1}(r) = 1 - \frac{2GM}{r}$.

We find the following solution to the eqs. of motion (and constraints) 
\rf{constr-0}--\rf{box-3}. Using the ansatz:
\be
X^0 \equiv t = \t \quad,\quad X^1 \equiv r = r(\t,\s^1,\s^2) \quad, \quad
X^2 \equiv \th = \th (\s^1,\s^2) \quad ,\quad  X^3 \equiv \p = \p (\s^1,\s^2) \; ,
\lab{ansatz-schwarzschild}
\ee
\be
\g_{ij} = a (\t)\, {\wti \g}_{ij} (\s^1,\s^2) \; ,
\lab{conf-flat}
\ee
with ${\wti \g}_{ij}$ being some standard reference $2D$ metric on the membrane
surface ($i,j=1,2$),
we obtain from Eqs.\rf{constr-0} taking into account \rf{spherical-symm-metric}:
\be
\partder{}{\t}r = \pm A(r) \quad ,\quad \partder{}{\s^i}r = 0 \; .
\lab{sol-schwarzschild-1}
\ee
From Eq.\rf{u-eqs-fix} we get $\partder{}{\t}r = 0$ which upon combining with 
\rf{sol-schwarzschild-1} gives:
\be
r = r_0 \equiv 2GM = \mathrm{const} \quad,\quad {i.e.}\;\; A(r_0)=0  \; .
\lab{sol-schwarzschild-2}
\ee
For the rest of embedding coordinates and the intrinsic {\em WILL}-membrane
metric (upon assuming the membrane surface to be of spherical topology) we obtain:
\be
\th = \s^1 \quad,\quad \p = \s^2  \quad,\quad 
\Vert\g_{ij}\Vert = c_0 \, e^{\mp \t/r_0}\,\twomat{1}{0}{0}{\sin^2 (\s^1)} \; ,
\lab{sol-schwarzschild-3}
\ee
where $c_0$ is an arbitrary integration constant.

That is, the {\em WILL}-membrane with spherical topology (and with
exponentially blowing-up/deflating internal metric) ``sits'' on (materializes) the
event horizon of the Schwarzschild black hole.


\subsection{Example: WILL-Membrane in a Reissner-Nordstr\"{o}m Black Hole}

Now we need to extend the {\em WILL}-brane model \rf{WI-brane} via a coupling 
to external space-time electromagnetic field $\cA_\m$. The natural
Weyl-conformal invariant candidate action reads (for $p=2$):
\be
S = - \int d^3\s \,\P (\vp) 
\Bigl\lb \h \g^{ab} \pa_a X^{\m} \pa_b X^{\n} G_{\m\n}
- \sqrt{F_{ab} F_{cd} \g^{ac}\g^{bd}}\Bigr\rb
- q\int d^3\s \, \vareps^{abc} \cA_\m \pa_a X^\m F_{bc} \; .
\lab{WILL-membrane+A}
\ee
The last Chern-Simmons-like term is a special case of a class of
Chern-Simmons-like couplings of extended objects to external electromagnetic
fields proposed in ref.\ct{Aaron-Eduardo}.

In the dual formulation we get accordingly:
\be
S_{\mathrm{dual}} = - \h \int d^3\s\, \chi (\g,u,\cA)\,\sqrt{-\g}
\g^{ab}\(\pa_a X \pa_b X\)  \; ,
\lab{WILL-membrane+A-dual}
\ee
with a variable brane tension:
\be
\chi (\g,u,\cA) \equiv \sqrt{-2\g^{cd}\(\pa_c u - q \cA_c\)
\(\pa_d u - q \cA_d\)} \quad ,\;\; \cA_a \equiv \cA_\m \pa_a X^\m \; .
\lab{tension+A}
\ee
Here $u$ is the dual ``gauge'' potential w.r.t. $A_a$ and the corresponding
field-strength and dual field-strength are related as:
\be
F_{ab}(A) = - \frac{1}{2\chi (\g,u,\cA)} \sqrt{-\g} \vareps_{abc} \g^{cd}
\(\pa_d u - q\cA_d \)\,\g^{ef}\!\(\pa_e X \pa_f X\)  \; .
\lab{a}
\ee
The extended {\em WILL}-membrane model in the dual formulation 
\rf{WILL-membrane+A-dual} is likewise manifestly Weyl-invariant 
(under $\g_{ab} \to \rho\g_{ab}$).

The eqs. of motion w.r.t. $\g^{ab}$, $u$ (or $A_a$), and $X^\m$ read
accordingly:
\be
\(\pa_a X \pa_b X\) + \h \g^{cd}\(\pa_c X \pa_d X\)
\(\frac{\(\pa_a u - q\cA_a\)\(\pa_b u -q\cA_b\) }
{\g^{ef} \(\pa_e u - q\cA_e\)\(\pa_f u - q\cA_f\)} - \g_{ab}\) = 0  \; ;
\lab{gamma-eqs+A}
\ee
\be
\pa_a \(\,\frac{\sqrt{-\g}\g^{ab}\(\pa_b u - q\cA_b\)}{\chi (\g,u,\cA)}\,
\g^{cd}\(\pa_c X \pa_d X\)\,\) = 0  \; ;
\lab{u-eqs+A}
\ee
\br
\pa_a \(\chi (\g,u,\cA)\,\sqrt{-\g} \g^{ab}\pa_b X^\m \) +
\chi (\g,u,\cA)\,\sqrt{-\g} \g^{ab}\pa_a X^\n \pa_b X^\l \G^\m_{\n\l} 
\nonu \\
- \; q \vareps^{abc} F_{bc} \pa_a X^\n 
\(\pa_\l \cA_\n - \pa_\n \cA_\l\)\, G^{\l\m} = 0  \; . \phantom{aaaaaaaa}
\lab{X-eqs+A}
\er

Using the same (synchronous) gauge choice \rf{gauge-fix} and 
ansatz for the dual ``gauge potential'' \rf{u-ansatz}, as well as
considering static external space-time electric field ($\cA_0 = Q/\sqrt{4\pi}\,r$ 
-- relevant case for Reissner-Nordstr\"{o}m blackholes, see next Section), the eqs. 
of motion w.r.t. $\g^{ab}$, $u$ (or $A_a$) and $X^\m$ acquire the form
(recall $\(\pa_a X \pa_b X\) \equiv \pa_a X^\m \pa_b X^\n G_{\m\n}$):
\be
\(\pa_0 X \pa_0 X\) = 0 \quad ,\quad \(\pa_0 X \pa_i X\) = 0  \; ,
\lab{constr-0+A}
\ee
\be
\(\pa_i X\pa_j X\) - \h \g_{ij} \g^{kl}\(\pa_k X\pa_l X\) = 0 \; ,
\lab{constr-vir+A}
\ee
(these constraints are the same as in the absence of coupling to space-time
gauge field \rf{constr-0}--\rf{constr-vir});
\be
\pa_0 \(\sqrt{\g_{(2)}} \g^{kl}\(\pa_k X\pa_l X\)\) = 0 \; ,
\lab{u-eqs-fix+A}
\ee
(once again the same equation as in the absence of coupling to space-time
gauge field \rf{u-eqs-fix});
\be
{\wti \Box}^{(3)} X^\m + \( - \pa_0 X^\n \pa_0 X^\l + 
\g^{kl} \pa_k X^\n \pa_l X^\l \) \G^{\m}_{\n\l} 
- q \frac{\g^{kl}\(\pa_k X \pa_l X\)}{\sqrt{2}\,\chi}\,
\pa_0 X^\n \(\pa_\l \cA_\n - \pa_\n \cA_\l\)\, G^{\l\m} = 0 \; ,
\lab{X-eqs-fix+A}
\ee
where $\chi \equiv T_0 - \sqrt{2}q\cA_0$ (the variable brane tension), 
$\cA_1 =\ldots = \cA_{D-1} =0$, and:
\be
{\wti \Box}^{(3)} \equiv 
- \frac{1}{\chi \sqrt{\g^{(2)}}} \pa_0 \(\chi \sqrt{\g^{(2)}} \pa_0 \) + 
\frac{1}{\chi \sqrt{\g^{(2)}}}\pa_i \(\chi \sqrt{\g^{(2)}} \g^{ij} \pa_j \)
\; .
\lab{box-3+A}
\ee

Now, let us solve Eqs.\rf{constr-0+A}--\rf{box-3+A} in Reissner-Nordstr\"{o}m 
background:
\be
(ds)^2 = - A(r)(dt)^2 + A^{-1}(dr)^2 + r^2 \lb (d\th)^2 + \sin^2 (\th)\,(d\p)^2\rb
\lab{spherical-symm-metric-a}
\ee
\be
A(r) = 1 - \frac{2GM}{r} + \frac{GQ^2}{r^2}.
\lab{RN-metric}
\ee
Employing the same ansatz \rf{ansatz-schwarzschild} as in the case of 
Schwarzschild background, the solution for Reissner-Nordstr\"{o}m background
reads:
\be
X^0 \equiv t = \t \quad,\quad
\th = \s^1 \quad,\quad \p = \s^2
\lab{RN-sol-1}
\ee
\be
r (\t,\s^1,\s^2) = r_{\mathrm{horizon}} = \mathrm{const}
\lab{RN-sol-2}
\ee
where $A(r_{\mathrm{horizon}}) = 0$;
\be
\Vert\g_{ij}\Vert = \( c_0 \, e^{\mp \t \(\partder{}{r}A\)_{r=r_{\mathrm{horizon}}}}
+ \frac{qQ}{\sqrt{2\pi}\(\chi\partder{}{r}A\)_{r=r_{\mathrm{horizon}}}}\)
\,\twomat{1}{0}{0}{\sin^2 (\s^1)}
\lab{RN-sol-3}
\ee
where $c_0$ is an arbitrary integration constant (recall 
$\chi \equiv T_0 - \sqrt{2}q\cA_0$).

In particular, taking $c_0 =0$ one obtains the usual time-independent internal 
spherical metric on the brane surface.
Thus, similar to the Schwarzschild case, the {\em WILL}-membrane with spherical 
topology ``sits'' on (materializes) the event horizon of the 
Reissner-Nordstr\"{o}m black hole.


\section{Coupled Einstein-Maxwell-{\em WILL}-Membrane System}

We can extend the results from the previous section to the case of the full
coupled Einstein-Maxwell-{\em WILL}-membrane system, \textsl{i.e.}, taking
into account the back-reaction of the {\em WILL}-membrane serving as a
material and electrically charged source for gravity and electromagnetism.
The pertinent action reads:
\be
S = \int\!\! d^4 x\,\sqrt{-G}\,\llb \frac{R}{16\pi G_N}
- \frac{1}{4} \cF_{\m\n}(\cA) \cF_{\k\l}(\cA) G^{\m\k} G^{\n\l}\rrb
+ S_{\mathrm{WILL-brane}}  \; ,
\lab{E-M-WILL} 
\ee
where $\cF_{\m\n}(\cA) = \pa_\m \cA_\n - \pa_\n \cA_\m$, and:
\br
S_{\mathrm{WILL-brane}} = - \int\!\! d^3\s \,\P (\vp) 
\Biggl\lb \h \g^{ab} \pa_a X^{\m} \pa_b X^{\n} G_{\m\n}
- \sqrt{F_{ab} F_{cd} \g^{ac}\g^{bd}}\,\Biggr\rb
- q\int\!\! d^3\s \, \vareps^{abc} \cA_\m \pa_a X^\m F_{bc}  \; .
\lab{WILL-membrane+A-a}
\er

Eqs. of motion for the \textsl{WILL}-membrane subsystem are the same as
above, namely Eqs.\rf{constr-0+A}--\rf{box-3+A}. The rest of the eqs. of motion are:
\be
R_{\m\n} - \h G_{\m\n} R = 8\pi G_N \( T^{(EM)}_{\m\n} + T^{(brane)}_{\m\n}\)\; ,
\lab{Einstein-eqs}
\ee
\be
\pa_\n \(\sqrt{-G}G^{\m\k}G^{\n\l} \cF_{\k\l}\) + j^\m = 0 \; ,
\lab{Maxwell-eqs}
\ee
where:
\be
T^{(EM)}_{\m\n} \equiv \cF_{\m\k}\cF_{\n\l} G^{\k\l} - G_{\m\n}\frac{1}{4}
\cF_{\r\k}\cF_{\s\l} G^{\r\s}G^{\k\l} \; ,
\lab{T-EM}
\ee
\be
T^{(brane)}_{\m\n} \equiv - G_{\m\k}G_{\n\l}
\int\!\! d^3 \s\, \frac{\d^{(4)}\Bigl(x-X(\s)\Bigr)}{\sqrt{-G}}\,
\chi\,\sqrt{-\g} \g^{ab}\pa_a X^\k \pa_b X^\l \; ,
\lab{T-brane}
\ee
(recall $\chi \equiv \sqrt{-2\g^{cd}\(\pa_c u - q \cA_c\)
\(\pa_d u - q \cA_d\)}$,  $\cA_a \equiv \cA_\m \pa_a X^\m$),
\be
j^\m \equiv q \int\!\! d^3 \s\,\d^{(4)}\Bigl(x-X(\s)\Bigr)
\vareps^{abc} F_{bc} \pa_a X^\m \; .
\lab{brane-EM-current}
\ee

Following the same steps as in the previous section we obtain the following 
spherically symmetric stationary solution. For the Einstein subsystem we
find a solution:
\be
(ds)^2 = - A(r)(dt)^2 + A^{-1}(dr)^2 + r^2 \lb (d\th)^2 + \sin^2 (\th)\,(d\p)^2\rb
\; ,
\lab{spherical-symm-metric-b}
\ee
consisting of two different black holes with a {\em common} event horizon:
\begin{itemize}
\item    
Schwarzschild black hole inside the horizon:
\be
A(r)\equiv A_{-}(r) = 1 - \frac{2GM_1}{r}\;\; ,\quad \mathrm{for}\;\; 
r < r_0 \equiv r_{\mathrm{horizon}}= 2GM_1 \; .
\lab{Schwarzschild-metric-in}
\ee
\item
Reissner-Norstr\"{o}m black hole outside the horizon:
\be
A(r)\equiv A_{+}(r) = 1 - \frac{2GM_2}{r} + \frac{GQ^2}{r^2}\;\; ,
\quad \mathrm{for}\;\; r > r_0 \equiv r_{\mathrm{horizon}} \; ,
\lab{RN-metric-out}
\ee
where $Q^2 = 8\pi q^2 r_{\mathrm{horizon}}^4 \equiv 128\pi q^2 G^4 M_1^4$;
\end{itemize}
For the Maxwell subsystem we get $\cA_1 = \ldots =\cA_{D-1}=0$ everywhere and:
\begin{itemize}
\item
Coulomb field outside horizon:
\be
\cA_0 = \frac{\sqrt{2}\, q\, r_{\mathrm{horizon}}^2}{r} \;\; ,\quad \mathrm{for}\;\; 
r \geq r_0 \equiv r_{\mathrm{horizon}}  \; .
\lab{EM-out}
\ee
\item
No electric field inside horizon:
\be
\cA_0 = \sqrt{2}\, q\, r_{\mathrm{horizon}} = \mathrm{const} \;\; ,\quad \mathrm{for}\;\; 
r \leq r_0 \equiv r_{\mathrm{horizon}}  \; .
\lab{EM-in}
\ee
\end{itemize}
The \textsl{WILL}-membrane again ``sits'' on (materializes) the common event 
horizon of the pertinent black holes:
\be
X^0 \equiv t = \t \quad,\quad \th = \s^1 \quad,\quad \p = \s^2 \quad,\quad 
r (\t,\s^1,\s^2) = r_{\mathrm{horizon}} = \mathrm{const}
\lab{Schwarzschild-RN-sol}
\ee
  
In addition there is an important matching condition for the metric components
along the \textsl{WILL}-membrane:
\be
\partder{}{r} A_{+}\bgv_{r=r_{\mathrm{horizon}}} -
\partder{}{r} A_{-}\bgv_{r=r_{\mathrm{horizon}}} = - 16\pi G \chi \; ,
\lab{metric-match}
\ee
which yields the following relations between the parameters of the black
holes and the \textsl{WILL}-membrane ($q$ being its surface charge density) :
\be
M_2 = M_1 + 32\pi q^2 G^3 M_1^3
\lab{mass-match}
\ee
and for the brane tension $\chi$:
\be
\chi \equiv T_0 - 2q^2r_{\mathrm{horizon}} =q^2 G M_1 \quad, \;\;
\mathrm{i.e.} \;\; T_0 = 5 q^2 G M_1
\lab{tension-match}
\ee

The matching condition \rf{metric-match} corresponds to the statically
soldering conditions in the light-like thin shell dynamics in general 
relativity \ct{Barrabes-Israel}. On the other hand we should stress that
unlike the latter phenomenological models of thin shell dynamics
(\textsl{i.e.}, where the membranes are introduced {\em ad hoc}), the present
{\em WILL}-brane models provide a systematic description of light-like branes
from first principles starting with concise Weyl-conformal invariant actions
\rf{WI-brane}, \rf{E-M-WILL}--\rf{WILL-membrane+A-a}. As a consequence, these
actions also yield additional information impossible to obtain within the
phenomenological approach, such as the requirement that the light-like brane
must sit on the event horizon of the pertinent black hole. 

\section{Conclusions and Outlook} 

In the present work we have demonstrated that employing alternative
non-Riemannian world-sheet/world-volume integration measure significantly 
affects string and $p$-brane dynamics:
\begin{itemize}
\item
Acceptable dynamics in the novel class of string/brane models
(Eqs.\rf{m-string} and \rf{WI-brane}) {\em naturally} requires the introduction
of auxiliary world-sheet/world-volume gauge fields.
\item
By employing square-root Yang-Mills actions for the auxiliary
world-sheet/world-volume gauge fields one achieves manifest {\em Weyl-conformal
symmetry} in the new class of $p$-brane theories {\em for any $p$}.
\item
The string/brane tension is {\em not} a constant dimensionful scale given 
{\em ad hoc}, but rather it appears as an {\em additional dynamical degree of 
freedom} beyond the ordinary string/brane degrees of freedom.
\item
The novel class of Weyl-invariant $p$-brane theories describes
intrinsically {\em light-like} $p$-branes for any even $p$ ({\em WILL}-branes).
\item
When put in a gravitational black hole background, the {\em WILL}-membrane
($p=2$) sits on (``materializes'') the event horizon.
\item
The coupled Einstein-Maxwell-{\em WILL}-membrane system \rf{E-M-WILL} possesses
self-consistent solution where the {\em WILL}-membrane serves as a 
material and electrically charged source for gravity and electromagnetism,
and it ``sits'' on (materializes) the common
event horizon for a Schwarzschild (in the interior) and Reissner-Nordstr\"{o}m
(in the exterior) black holes. Thus our model \rf{E-M-WILL} provides an
explicit dynamical realization of the so called ``membrane paradigm'' in the
physics of black holes \ct{membrane-paradigm}. 
\end{itemize}

One can think of various physically interesting directions of further
reseach on the novel class of Weyl-conformal invariant $p$-branes such as:
quantization (Weyl-conformal anomaly and critical dimensions); supersymmetric
extension; possible relevance for the open string dynamics (similar to the
Dirichlet- ($Dp$-)branes); {\em WILL}-brane dynamics in more complicated
gravitational black hole backgrounds (\textsl{e.g.}, Kerr-Newman) \textsl{etc.}.

\vspace{.1in}

\textbf{Acknowledgements.}
{\small Two of us (E.N. and S.P.) are especially grateful
to Prof. Branko Dragovich for his cordial hospitality at the Third Summer
School on Modern Mathematical Physics, Zlatibor (Serbia and Montenegro), 2004,
where the above results were first reported. They are also sincerely thankful to 
the organizers for hospitality and support at the 2nd Annual Meeting of the European
RTN \textsl{EUCLID}, Sozopol (Bulgaria), 2004. One of us (E.G.) thanks Dr.
Stefano Ansoldi for useful conversations concerning light-like shell
dynamics and for pointing out to us the treatment in Ref.\ct{Barrabes-Israel}. 

E.N. and S.P. are partially supported by Bulgarian NSF grants \textsl{F-904/99}
and \textsl{F-1412/04}.
Finally, all of us acknowledge support of our collaboration through the exchange
agreement between the Ben-Gurion Univesity of the Negev and the Bulgarian
Academy of Sciences.}

\end{document}